\documentclass[aps,prl,nofootinbib,square,sort&compress,10pt,fleqn,showpacs,longbibliography]{revtex4-2}
\usepackage[english]{babel}
\usepackage{amssymb}
\usepackage{upgreek}
\usepackage{comment}
\usepackage{yfonts}
\usepackage[fleqn]{amsmath}
\usepackage[T1]{fontenc}
\usepackage[x11names]{xcolor}
\usepackage{empheq}
\usepackage{color}
\allowdisplaybreaks
\usepackage{mathtools}
\usepackage{eucal}
\usepackage{graphicx}
\usepackage{float}
\usepackage[pdfencoding=auto,psdextra]{hyperref}
\usepackage{bookmark}
\usepackage{url}
\usepackage{epsfig}
\usepackage{bm}
\usepackage[utf8x,latin1]{inputenc} 
\usepackage{emerald}
\usepackage{array}
\let\tempcal \mathcal
\usepackage{dutchcal}

\let\mathcal\tempcal

\makeatletter
\newcommand{\mathleft}{\@fleqntrue\@mathmargin0pt}
\newcommand{\mathcenter}{\@fleqnfalse}

\DeclareMathAlphabet{\mathpzc}{OT1}{pzc}{m}{it}

\def\be{\begin{equation}}
\def\ee{\end{equation}}
\def\ba{\begin{eqnarray}}
\def\bal{\begin{eqnarray}\label}
\def\ea{\end{eqnarray}}

\def\M{\mathcal{M}}

\def\V{\mathcal{V}}

\def\x{{\bm x}}
\def\X{{\bm X}}

\def\g{\mathfrak{g}}
\def\L{{\pounds}}

\def\vp{\varphi}

\def\pd{\partial}

\begin{document}
\title{Revolutionizing Gravitational Potential Analysis: From Clairaut to Lie Groups}
\author{Sergei M. Kopeikin}
\affiliation{Department of Physics \& Astronomy, University of Missouri, 322 Physics Bldg., Columbia, Missouri 65211, USA}
\email{E-mail: kopeikins@missouri.edu}
\date{\today}

\begin{abstract}\noindent
This letter introduces an advanced novel theory for calculating non-linear Newtonian hydrostatic perturbations in the density, shape, and gravitational field of fluid stars and planets subjected to external tidal and rotational forces. The theory employs a Lie group approach using exponential mappings to derive exact differential equations for large gravitational field perturbations and the shape function, which describes the finite deformation of the body's figure. This approach lays the foundation for the precise analytic determination and numerical computation of the induced body's multipole moments and Love numbers with any desired degree of accuracy.
\end{abstract}
\pacs{}
\maketitle

\section{Introduction}\label{sec-1}
The exact calculation of the elastic response of a celestial fluid body to external perturbations caused by tidal gravitational and/or rotational forces has been a formidable task for generations of mathematical physicists and astrophysicists. A fluid body occupies a volume whose boundary is an equipotential level surface of gravitational potential. Thus, the problem is to determine the distortion in the shape of the level surfaces induced by the external gravitational force. Despite its seemingly simple appearance, the problem involves multiple serious challenges. 

The initial breakthroughs in addressing this problem were achieved by Clairaut \citep{Clairaut-book-1743} and, then, by Darwin \citep{Darwin_1899} and de Sitter \citep{deSitter_1924BAN}. Successive progress was made by a number of distinguished scientists, including Love, Poincar\'e, Lyapunov, Chandrasekhar, de Sitter, Kopal, Molodensky, Zharkov, Hubbard, among others (refer to historical reviews in publications \citep{chandr87, Efimov_1978,Hubbard-book,Jardetzky_2005, horedt_2004book}). The most commonly employed method for determining the shape of a rotating and tidally perturbed astronomical body involves the Legendre series decomposition of the gravitational potential into spherical harmonics. The requirement that the potential remains constant on the level surface results in an infinite set of coupled, non-linear integro-differential equations for the amplitudes of each spherical harmonic. These equations are subsequently solved using the method of successive iterations. This process is exceedingly laborious, as the complexity of the equations increases rapidly with higher-order harmonics. Consequently, these equations have only been resolved up to a limited order of approximation \citep{Nettelmann_2021}. Further advancements in the application of the Legendre decomposition method are uncertain, necessitating the development of a more sophisticated theory for the non-linear perturbations of the shapes of rotating stars and planets \citep{Hubbard_2013}.  

The recent advent of gravitational wave astronomy provides an additional compelling impetus for developing a non-linear theory of gravitational perturbations in astronomical bodies. The collision of neutron stars within a binary system, driven by the emission of gravitational waves, offers a unique opportunity to investigate the equation of state of matter at nuclear densities. The tidal deformations of these stars are quantified by Love numbers \citep{Love_1909}, whose calculation requires a thorough understanding of the non-linear strong field regime in both Newtonian gravity and general relativity. Although modern computational technologies can produce numerical solutions to this problem, they still lack the validation that exact analytic theory provides, offering an independent and robust confirmation of the numerical results. Consequently, the development of a non-linear analytic theory of Love numbers is a crucial aspect of contemporary gravitational wave astronomy \citep{Yip_2017}.

This letter marks a pivotal advancement in the development of an exact non-linear theory of hydrostatic Newtonian tidal and rotational perturbations for an idealized yet realistic model of stars and planets composed of an ideal fluid. It extends the most sophisticated analytic theory of these perturbations, originally formulated by Zharkov \citep{Zharkov-book-1986}, to encompass arbitrary higher orders of approximation. This enhanced theory is also applicable for calculating the Love numbers and internal structure of rapidly rotating giant planets in the solar system, such as Jupiter and Saturn, with any desired degree of precision.

The theoretical framework established in this letter employs an innovative approach based on the Lie group of diffeomorphisms, $G=\text{Diff}(\M)$ for fluid dynamics. This approach was pioneered by V. Arnold \cite{arnold} and further developed by his disciples \citep{khesin}. The base manifold $\M$ of the group is the three-dimensional volume ${\cal V}$ of the celestial body (star or planet) filled with an ideal fluid. In the absence of perturbation the volume $\V$ is a sphere with a radial density profile, $\rho:=\rho(r)$. External perturbations (rotation, tide) are described by gravitational potential $W$. These perturbations cause each element of the fluid to change its unperturbed position $\x\in\M$ to a new position $\x_\tau\in\M_\tau$, deforming the base manifold $\M\to\M_\tau$ and the body's volume $\V\to\V_\tau$ where $\tau\in[0,1]$ is a parameter characterizing the magnitude of the deformation.  The map, $\vp_\tau:~\x\to\x_\tau$, is a one-parameter diffeomorphism known as a vector flow. The set of all vector flows forms the Lie group $G$. Infinitesimally small diffeomorphisms are linear with respect to the parameter $\tau$ and form the Lie algebra $\g= \text{Vect}(\M)$ under the Lie bracket operation. The finite, non-linear response of the fluid body to the external perturbation is described by the exponential map of the elements of the Lie algebra to the Lie group. 

The objective of the present perturbation theory is twofold: 1) to analytically describe the non-linear (hydrostatic) response of the body's density and gravitational potential to presumably strong, external perturbations characterized by the potential $W$, and 2) to determine the shape of the strongly perturbed body's volume ${\cal V}_\tau$ in the limit $\tau\to 1$. This is achieved using the sophisticated mathematical framework of Lie groups of vector flows \citep{khesin}. Initially, we construct the Lie algebra $\mathfrak{g}$ of the group and derive the equations of hydrostatic equilibrium for the fluid and the gravitational field on $\mathfrak{g}$. Subsequently, we extend the Lie algebra equations to the Lie group by applying the formalism of the exponential map of the vector flow. Ultimately, we derive exact (non-linear) differential equations for the shape function, gravitational field, and density perturbations of the body, and express the induced multipole moments of the body in terms of the shape function.

We use $x^i$ for the Cartesian coordinates covering the entire space $\mathbb{R}^3$. The spherical coordinates are denoted $(r,\theta,\phi)$. The unit vector along the radius is ${\bm n}=(n^i)=x^i/r$. Boldface letters are used to denote spatial vectors, e.g., $\x=(x^i)$, ${\bm\xi}=(\xi^i)$, etc. The origin of the coordinates is at the center of mass of the body. Spatial indices $i,j,k$ take values $1,2,3$. Partial derivatives are denoted ${\bm\nabla}=(\pd_i)\equiv\pd/\pd x^i$. Multi-index derivative are denoted $\pd_{i_1\ldots i_n}=\pd^n/\pd x^{i_1}\ldots\pd x^{i_n}$. A prime denotes a partial derivative with respect to the radial coordinate, $f'=\pd_r f$. The Kronecker symbol $\delta^{ij}={\rm diag}(1,1,1)$.  Repeated spatial indices denote the Einstein summation rule: $a^ib_i=a^1b_1+a^2b_2+a^3b_3$. In what follows, we assume that all functions involved in the analysis are analytic and that all infinite sums are convergent. We also use geometric units with the universal gravitational constant $G=1$. 

\section{The Base Manifold}\label{sec-4}

The base manifold $\M\in\mathbb{R}^3$ of the Lie group $G$ represents the interior of a celestial body of mass $M$ consisting of an {\it ideal} fluid. In the absence of external perturbations $\M$ is a sphere with a constant radius, normalized to $r=1$. The manifold $\M$ is populated with scalar functions that characterize the fluid: density $\rho$, pressure $p$, and gravitational potential of the body $U$.

In this letter, we assume that pressure depends on density through a barotopic equation of state $p=p(\rho)$, while the gravitational potential satisfies the Poisson equation:
\bal{3}
\Delta U=-4\pi\rho\;,
\ea
where $\Delta=\delta^{ij}\pd_{ij}$ is the Laplacian. Eq. \eqref{3} has the solution:
\bal{b2}
U(\x)&=&\int_{\V}d^3x'\frac{\rho(\x')}{|\x-\x'|}\;.
\ea
The fluid is in hydrostatic equilibrium described by equation:
\bal{4}
\rho\pd_i U&=&\pd_i p\;.
\ea
In the absence of external perturbations all functions residing on the base manifold $\M$ depend only on the radial coordinate, $\rho=\rho(r)$, $p=p(\rho(r))$, $U=U(r)$. Consequently, the gradient of any of these functions is directed along the radius, e.g., $\pd_i\rho=n^i\rho'$. We assume that the solution of Eqs. \eqref{3}--\eqref{4} on the base manifold $\M$ is known.

\section{The Lie Group of Vector Flows}\label{sec-2}

Let us consider a perturbation of the fluid body cause by an external gravitational potential $W$ that obeys the Laplace equation:
\bal{1c}
\Delta W&=&0\;.
\ea 
We assume the response of the body to the perturbation is instantaneous so that time-dependent terms are neglected.

Due to the external perturbation, each element of the fluid is displaced in $\mathbb{R}^3$ by a finite distance along the integral curve, $\x_\tau=\x(\tau)$, of the vector flow, which is governed by a first-order differential equation:
\bal{6}
\frac{dx^i_\tau}{d\tau}&=&\xi^i(\x_\tau)\;,
\ea
where vector $\xi^i$ is called a generator of the flow. The set of all generators $\xi^i=\xi^i(\x)$ forms the Lie algebra $\g$ of the group $G$. Vector field ${\bm\xi}$ is smooth but is not subject to any other additional constraints. The solution of Eq. \eqref{6} is given in the neighborhood of point $\x=\x_0$ by the push-forward exponential map:
\bal{7}
x^i_\tau&=&\exp\left(\tau L_{\bm\xi}\right)x^i=x^i+\tau\xi^i+\frac{\tau^2}{2!}\xi^j\pd_j\xi^i+\frac{\tau^2}{3!}\xi^p\pd_p\left(\xi^j\pd_j\xi^i\right)+...\;, 
\ea
where all vectors and their derivatives refer to the point $\x\in\M$, and $L_{\bm\xi}\equiv \xi^i\pd_i$ is the operator of directional derivative along the vector $\xi^i=\xi^i(\x)$. The point $\x=\x_0$ represents the identity element of the flow.  

The vector flow forms a one-parametric group of diffeomorphisms $\phi_\tau:~\x\to\x_\tau$. The inverse, or pull back element of the flow $x^i_{-\tau}$, is
obtained from Eq. \eqref{7} by mapping $\tau\to -\tau$.
It is straightforward to check that $\x_{-\tau}\cdot \x_{\tau}=\x_{\tau}\cdot\x_{-\tau}=\x$. The set of all vectors flow on $\M$ forms the group of diffeomorphisms $G=\text{Diff}(\M)$. The group $G$ is not compact and its space is dimensionless \citep{khesin}. 

In addition to the vector flows, there are various smooth functions $f$ on the manifold $\M$, such as fluid's density $\rho$, pressure $p$, gravitational potential $U$. These functions form a vector space $C^\infty$, which is incorporated into the Lie group $G$, extending it to a Lie groupoid $G \rightrightarrows \M$ \citep{Mackenzie_2005}.

The Lie algebra $\g$ of the Lie group $G$ is defined as the tangent space $\text{Vect}(\M)$ at the identity element $\text{\bf id}_\M:~\x\to\x$, which maps each point of the manifold $\M$ to itself. This Lie algebra is equipped with a bilinear operation called the Lie bracket or commutator. The commutator of two vector fields ${\bm\xi}\in \g$ and ${\bm\eta}\in \g$ is given by their Lie derivative: $[{\bm\xi},{\bm\eta}]=\L_{\bm\xi}{\bm\eta}=-\L_{\bm\eta}{\bm\xi}$. The commutator satisfies the Jacobi identity \citep{arnold}.

Compatibility of differential operations on functions $f\in C^\infty$ with the structure of the Lie group of diffeomorphisms $G$ is provided by the {\it anchor map} which extends the Lie algebra $\g$ to the Lie algebroid $\g\rightarrow \M$. The anchor map is defined 
as follows \cite{Mackenzie_2005}: 
\bal{1}
[{\bm\xi},f{\bm\eta}]&=&f[{\bm\xi},{\bm\eta}]+\pounds_{\bm\xi}f\cdot{\bm\eta}\;,
\ea
where $\pounds_{\bm\xi}$ is a Lie derivative of $f$ along the vector field ${\bm\xi}$. The anchor map extends the the concept of the commutator of two vector fields to the commutator of a vector field and a scalar function:
\bal{2}
[{\bm\xi},f]=-[f,{\bm\xi}]=\pounds_{\bm\xi}f\;.
\ea
Direct inspection confirms that definition \eqref{2} satisfies the Jacobi identity.

\section{Variations of Density and Gravitational Field}

The perturbation $W$ alters the body's volume $\V\to\V_\tau$ and causes all variables to change along the vector flow: $\rho\to\rho_\tau$, $p\to p_\tau$, $U\to U_\tau$. The perturbed gravitational field satisfies the Poisson equation:
\bal{3a}
\Delta U_\tau=-4\pi\rho_\tau\;,
\ea
and the equation of hydrostatic equilibrium:
\bal{4a}
\rho_\tau\pd_i\left(U_\tau+\tau W\right)&=&\pd_i p_\tau\;,
\ea
where the perturbed pressure $p_\tau$  is governed by the known equation of state, $p_\tau=p\left(\rho_\tau\right)$. The external perturbation $W$ obeys the Laplace equation \eqref{1c} and is not subject to variation. 

The perturbations of density and pressure are defined on the manifold $\M$ as the Eulerian variations induced by the vector flow $\x_\tau$ \citep{chandr87}. All variations are functions of unperturbed coordinates $\x$. Specifically, the generator $\xi^i=\xi^i(\x)$ of the Lie algebra of the vector flow at the point $\x\in{\cal M}$ gives rise to the first-order perturbation of density, $\delta_{\bm\xi}\rho$, on the tangent bundle $T_\x\M$ of the base manifold $\M$, which is naturally defined in terms of the Lie derivative along the vector ${\bm\xi}$ \citep{chandr87}:
\bal{aa1}
\delta_{\bm\xi}\rho&=&\pounds_{\bm\xi}\rho\;.
\ea 
The density is a scalar of weight $-1$. Therefore, the Lie derivative of the density is given by \citep{arnold}: 
\bal{nb6}
\pounds_{\bm\xi}\rho&=&-\pd_i\left(\rho\xi^i\right)=-\xi^i\pd_i\rho-\rho\theta\;,
\ea 
where $\theta=\pd_i\xi^i$ is a divergence of the vector flow.
The first-order perturbation of pressure $p=p(\rho)$ is obtained by taking the Lie derivative from the equation of state:
\bal{n7g}
\delta_{\bm\xi}p&=&\frac{\pd p}{\pd\rho}\delta_{\bm\xi}\rho\;.
\ea
The gravitational field $U$ is a functional of the density: $U=U[\rho]$ as shown in Eq. \eqref{b2}. Therefore, the linearized perturbation of the gravitational field $U$ is given by the variational derivative of the functional with respect to the density \citep{chandr87}:
\bal{oi7}
\delta_{\bm\xi}U&=&U[\delta_{\bm\xi}\rho]=\int_{\V}d^3x'\frac{\delta_{\bm\xi}\rho(\x')}{|\x-\x'|}\;,
\ea
where the integration is over the undisturbed volume $\V$.

The extension of the first-order perturbations from the $T\M$ to the finite perturbations on the entire manifold $T\M\to\M$ is achieved by the exponential mapping of the elements of the Lie algebra to the elements of the Lie group \citep{khesin,dfn}:
\bal{aa3}
\rho_\tau(\x)&=&\exp\left(\tau\delta_{\bm\xi}\right)\rho(\x)\;,\\
\label{vtf6}
U_\tau(\x)&=&\int_{\V}d^3y\frac{\rho_\tau(\x')}{|\x-\x'|}\;,
\ea
where the integration is over the undisturbed volume $\V$ of the base manifold $\M$.

Eq. \eqref{vtf6} is compatible with the perturbed field equation \eqref{3a} whose particular solution is given by
\bal{my3}
U^\dagger_\tau(\x)&=&\int_{\V_\tau}d^3x'_\tau\frac{\rho_\tau(\x'_\tau)}{|\x-\x'_\tau|}\;,
\ea
where the integration is carried out over the perturbed volume $\V_\tau$  of the body in coordinates $\x'_\tau\in\M_\tau$. By pulling back the coordinates $\x'_\tau\to\x'\in\M$ in the integral \eqref{my3}, the volume $\V_\tau$ transforms to $\V$. After a rather sophisticated calculation of the determinant of the pullback transformation and using the {\it anchor map} \eqref{2} to compute commutators of the vector fields and functions, the integrand reduces to the exponential Lie derivative:
\bal{po4c}
U^\dagger_\tau(\x)&=&\int_{\V}d^3x'\exp\left(-\tau\pounds_{{\bm\xi}'}\right)\left[\frac{\rho_\tau(\x')}{|\x-\x'|}\right]\;, 
\ea
where the integration goes over the unperturbed volume $\V\in\M$.
Here, the integral can be decomposed in two terms
\bal{1h}
U^\dagger_\tau(\x)&=&U_\tau(\x)+\oint_{\cal S}dS'_i\frac{\exp\left(-\tau\pounds_{{\bm\xi}'}\right)-1}{ \pounds_{{\bm\xi}'}}\left[\frac{\rho_\tau(\x')\xi'^i}{|\x-\x'|}\right]\;,
\ea
where $U_\tau$ is given in Eq. \eqref{vtf6}, and the surface integral is over the body's boundary, representing a solution of the homogeneous Laplace equation. In this letter, we assume that the density and all its derivatives vanish on the boundary surface. This assumption eliminates the surface integral, resulting in $U^\dagger_\tau(\x)=U_\tau(\x)$, q.e.d. 

It is convenient to introduce the total Eulerian variations of density and gravitational field, defined as follows:
\bal{9}
\varrho_\tau&\equiv&\rho_\tau(\x)-\rho(\x)\;,\\\label{9x}
V_\tau&\equiv&U_\tau(\x)-U(\x)\;
\ea
where $\rho_\tau$ and $U_\tau$ are defined in Eqs. \eqref{aa3} and \eqref{vtf6} respectively.
Let us also introduce the overall perturbation of the gravitational field:
\bal{b4}
K_\tau&\equiv&V_\tau+\tau W=\left[\exp\left(\tau\delta_{{\bm\xi}}\right)-1\right]U+\tau W\;.
\ea
Notice that the external perturbation $W$ is considered a known function that is not subject to variation \citep{chandr87}.

\section{Lie Group Techniques for Solving Hydrostatic Equilibrium Equations}\label{sec-5}

The equation of hydrostatic equilibrium \eqref{4a} establishes a differential relationship between perturbations in the gravitational field and density. This differential equation can be effectively solved using Lie group theory techniques.

First, let us consider Eq. \eqref{4a} for the elements of the Lie algebroid $\g\to{\cal M}$. By taking the partial derivative of Eq. \eqref{4a} with respect to $\tau$ and neglecting all higher-order terms in $\tau$, we obtain:
\bal{b1}
\delta_{{\bm\xi}}\rho\pd_i U+\rho\pd_i\left(\delta_{{\bm\xi}} U+W\right)=\pd_i(\delta_{{\bm\xi}}p)\;,
\ea 
where we have used the fact that the Eulerian variation and partial derivative commute, i.e., $[\delta_{{\bm\xi}},\pd_i]=0$. 
Next, we replace the first term on the left-hand side of Eq. \eqref{b1} with the undisturbed equation \eqref{4} and apply the following transformation:
\bal{b3}
\delta_{{\bm\xi}}\rho\pd_i U&=&\frac{1}{\rho}\pd_ip\delta_{{\bm\xi}}\rho=\frac{1}{\rho}\frac{\pd p}{\pd\rho}\pd_i\rho\delta_{{\bm\xi}}\rho=\frac{1}{\rho}\delta_{{\bm\xi}} p\pd_i\rho=\pd_i(\delta_{{\bm\xi}}p)-\rho\pd_i\left(\frac{\delta_{{\bm\xi}}p}{\rho}\right)\;.
\ea
Substituting Eq. \eqref{b3} back into Eq. \eqref{b1} yields:
\bal{4r}
\rho\pd_i\left(\delta_{{\bm\xi}} U+W-\rho^{-1}\delta_{{\bm\xi}}p\right)&=&0\;.
\ea
Integrating this equation establishes the correspondence between the elements of the Lie algebroid and the external perturbation: 
\bal{4s}
\delta_{{\bm\xi}}U+W&=&A(\rho)\delta_{{\bm\xi}}\rho\;,
\ea
where the function $A\equiv A(\rho)= \rho^{-1}\pd p/\pd\rho$.

Extention of Eq. \eqref{4s} to the elements of the Lie groupoid $G \rightrightarrows \M$ is achieved by applying the exponential mapping. This yields:
\bal{b5}
K_\tau&=&\frac{\exp\left(\tau\delta_{{\bm\xi}}\right)-1}{ \delta_{{\bm\xi}}}\left[A(\rho)\delta_{{\bm\xi}}\rho\right]\;,
\ea
where $K_\tau$ is given in Eq. \eqref{b4}. The right hand side of Eq. \eqref{b5} can be calculated by expanding the exponential operator into a Taylor series and applying the Fa\`a di Bruno formula \citep{deBruno}. Additional transformations, which are not shown here due to their technical complexity and cumbersome nature, yield:
\bal{b6}
K_\tau&=&\sum_{n=0}^\infty\frac{\varrho_\tau^{n+1}}{(n+1)!}\frac{\pd^nA(\rho)}{\pd\rho^n}\;,
\ea
where $\pd_\rho$ denotes the operator of a partial derivative with respective to density.
Substituting this expression for $K_\tau$ into Eq. \eqref{5} yields a differential equation for the density perturbation $\varrho_\tau$. However, it is more interesting to invert formula \eqref{b6} and derive the equation for the perturbation of the gravitational field $K_\tau$.

The inversion of Eq. \eqref{b6} is achieved using the Lagrange inversion formula \citep{lagrangeinv}, which gives:
\bal{s9}
\varrho_\tau&=&\frac{K_\tau}{A}+\sum_{n=2}^\infty\frac{g_{n}}{{n}!}\left(\frac{K_\tau}{A}\right)^{n}\;,
\ea
where coefficients
\bal{s10}
g_{n}&=&\sum_{k=1}^{{n}-1}(-1)^k({n})_{k}{\bf B}_{{n}-1,k}\left(\frac{\pd_\rho A}{2A},\frac{\pd_\rho^2 A}{3A},\dots,
\frac{1}{n-k+1}\frac{\pd_\rho^{n-k} A}{A}\right)\;,
\ea
are expressed in terms of the incomplete Bell polynomials ${\bf B}_{n,k}$ \citep{bellpol}, and $({n})_k:={n}({n}+1)...({n}+k-1)$
is the Pochhammer symbol. 

Notice that $A$ and its partial derivatives with respect to density can be reformulated in terms of the derivatives of the potential $U=U(r)$ and density $\rho=\rho(r)$. This is possible because the unperturbed density $\rho$ and the radial coordinate $r$ are bijective: $\rho\cong r$. Thus, the partial derivative $\pd_\rho=(1/\rho')\pd_r$ and Eq. \eqref{4} allow us to write the expression for function $A=A[\rho(r)]$ as follows: 
\bal{f8}
A&=&\frac{\pd U}{\pd\rho}=\frac{U'}{\rho'}\;,
\ea
where the prime denotes a derivative with respect to the radial coordinate $r$.  

\section{Equation for Gravitational Field Perturbation Decoupled from Density Variation}
The equation for gravitational field perturbation $K_\tau$ is obtained by taking the Laplacian of both sides of Eq. \eqref{b4} and incorporating Eqs. \eqref{3}, \eqref{1c} and \eqref{3a}:
\bal{5}
\Delta K_\tau&=&-4\pi\varrho_\tau\;.
\ea
This equation is valid inside the body. Outside the body, Eq. \eqref{5} is reduced to the Laplace equation, which is also valid inside the body with a homogeneous density distribution, where $\varrho_\tau=0$. Although Eq. \eqref{5} contains the density variation $\varrho_\tau$, it is expressed in terms of $K_\tau$ according to Eq. \eqref{s9}. Thus, substituting Eq. \eqref{s9} into Eq. \eqref{5} decouples perturbation of the gravitational field from the density variation and yields:
\bal{s14}
\Delta K_\tau+\kappa^2 K_\tau+4\pi\sum_{n=2}^\infty\frac{g_{n}}{{n}!}\left(\frac{K_\tau}{A}\right)^{n}&=&0\;,
\ea
where the coefficient $\kappa^2=4\pi/A$. 

At this point, it is worthwhile to note that the perturbations in density $\varrho_\tau$ and the gravitational field $K_\tau$ are governed by the physical deformation of the body's figure, which is determined by the vector flow $\x_\tau$ with generator ${\bm\xi}$. Interestingly, Eq. \eqref{s14} indicates that the perturbation of the gravitational field, $K_\tau$, can be determined without explicitly specifying the diffeomorphism ${\bm\xi}$. This is due to a specific gauge freedom inherent in the problem, defined by the transformation ${\bm\xi}\to{\bm\xi}+{\bm\chi}_\perp$, where the gauge vector field ${\bm\chi}_\perp$ is orthogonal to the radial direction, ${\bm n} \cdot {\bm\chi}_\perp = 0$, and volume-preserving, $\nabla \cdot {\bm\chi}_\perp = 0$. The addition of the field ${\bm\chi}_\perp$ does not affect the variation of density $\varrho_\tau$ or the gravitational field $K_\tau$. Hence, hydrostatic perturbations in an ideal fluid exhibit two unconstrained degrees of freedom.

Equation \eqref{s14} is valid within the volume occupied by the fluid body. If we omit the non-linear terms in Eq. \eqref{s14}, it reduces to a Helmholtz equation for the perturbation of the gravitational field, as first derived by S. Molodensky \citep{molodensky1953}. The formalism of Lie groups enables the extension of the Molodensky equation to the non-linear regime with arbitrary accuracy, revealing that gravitational field perturbations are self-interacting. At first glance, this may seem contradictory, as we are accustomed to thinking of the Newtonian field as linear and obeying the principle of superposition.

This is indeed true in a vacuum, where the Newtonian field is governed by the Laplace equation. However, inside matter, the gravitational field perturbs the matter density. The density response \eqref{s9} to the perturbation is linear only in the weak-field approximation. It becomes non-linear as the strength of the perturbation increases. This phenomenon is similar to what occurs in Maxwell's theory, where a strong electromagnetic perturbation causes non-linear polarization of matter, resulting in the Maxwell equations assuming a nonlinear form \citep{Ida_1997}. The self-interaction of the gravitational field perturbation $K_\tau$ exists only inside matter. In a vacuum, outside the body, the density and pressure vanish making coefficient $1/A\to 0$. Consequently, outside the body Eq. \eqref{s14} simplifies to the Laplace equation.

\section{Geometry and Functional Equations of Level Surfaces}\label{sec-6}

The level surface is defined as a surface of equal gravitational potential. For an ideal fluid, this surface also maintains constant values of density and pressure \citep{MacMillan1930}. The boundary of the body is the level surface where the pressure condition $p=0$ holds. We assume that the body's boundary has no surface density layer, so the fluid's density and all its derivatives vanish at the boundary as well. The level surfaces of the undisturbed fluid body are spherical, parameterized by the radial coordinate $r\in\M$. The external potential $W$ distorts these spherical level surfaces, rendering them non-spherical and mapping the base manifold $\M$ to $\M_\tau$. 

To determine an infinitesimal generator of the diffeomorphism that transforms an unperturbed to a perturbed level surface on the manifold $\M_\tau$, we must consider the gauge freedom ${\bm\xi}\to{\bm\xi}+{\bm\chi}_\perp$ discussed earlier. The physically-meaningful vector ${\bm\xi}$ is constrained only by the field equation \eqref{s14}, allowing two components of ${\bm\xi}$ to be chosen arbitrarily. To simplify calculations, it is convenient to choose the gauge vector ${\bm\chi}_\perp$ such that it eliminates two non-radial components of ${\bm\xi}$. We call this choice a radial gauge. In this gauge, we set ${\bm\xi} = \xi {\bm n}$ where $\xi=\xi(\x)$ depends on all three coordinates.

The radial gauge facilitates the establishment of functional equations that relate Eulerian variations to the magnitude of geometric deformation of level surfaces. To achieve this, we rewrite Eq. \eqref{b6} in terms of the potential $U=U(r(\rho))$ by using Eq. \eqref{f8}:
\bal{tt34}
K_\tau&=&\sum_{n=1}^\infty\frac{\varrho_\tau^{n}}{n!}\left(\frac{\pd U}{\pd \rho}\right)^n=\left[\exp\left(\tau\pounds_{\bm\xi}\rho\frac{\pd}{\pd\rho}\right)-1\right]U\;,
\ea
where the second term is obtained by accounting for the exponential form of $\varrho_\tau$ as defined in Eqs. \eqref{vtf6}, \eqref{9}, and the definition of the generating function of the Bell polynomials \citep{bellpol}. Because the derivative $\pd_\rho=(1/\rho')\pd_r$ Eq. \eqref{tt34} can be reformulated as an exponential (pull back) diffeomorphism:
\bal{bb7}
K_\tau&=&\bigl[\exp\left(-\tau L_{\bm\zeta}\right)-1\bigr]U(\x)\;,
\ea
where the linear operator $L_{\bm\zeta}\equiv\zeta\pd_r$, and the
generator $\zeta$ is expressed in terms of $\xi$ by the following equation:
\bal{qq7}
\zeta&=&-\frac{\pounds_{\bm\xi}\rho}{\rho'}=\xi+\frac{\rho}{\rho'}\left(\xi'+\frac{2\xi}{r}\right)\;.
\ea

The operator $L_{\bm\zeta}$ defines an infinitesimally small radial translation, but our goal is to derive an equation that allows us to determine large, finite deformations of the level surfaces. Therefore, it is more instructive to work directly with the translations describing the finite deformations of the fluid body. Each point $\x$ on the undisturbed spherical level surface of radius $r=|\x|$ is mapped by means of the radial diffeomorphism to a point $\x_\tau$ on the disturbed level surface: 
\bal{pa3}
x^i_\tau=x^i+X^i_\tau\;,
\ea 
where the vector of finite radial translation $X_\tau^i=X_\tau n^i$. The radial displacement of the level surface $X_\tau=X_\tau(\x)$ is called the height function \citep{Zharkov-book-1986}. The height function $X_\tau$ relates to the generator $\zeta$ by the (push forward) exponential map:
\bal{wx4}
X_\tau&=&\bigl[\exp\left(\tau L_{\bm\zeta}\right)-1\bigr]r=\tau\zeta+\frac{\tau^2}{2!}\zeta\pd_r\zeta+\frac{\tau^2}{3!}\zeta\pd_r\left(\zeta\pd_r\zeta\right)+\ldots\;. 
\ea
Using the vector $X^i_\tau$ allows us to write down Eqs. \eqref{bb7}, \eqref{qq7}, defining the finite deformations of the level surfaces in terms of the height function as push forward translations:
\bal{pi6c}
U(\x+\X_\tau)+K_\tau(\x+\X_\tau)&=&U(\x)\;,\\\label{pic7}
\rho(\x+\X_\tau)+\varrho_\tau(\x+\X_\tau)&=&\rho(\x)\;.
\ea

The next step is to solve the functional equations \eqref{pi6c}, \eqref{pic7} with respect to the height function $X_\tau$.  It is achieved by transforming Eq. \eqref{pi6c} into a fundamental differential equation for the height function $X$ in astronomical bodies. From this point forward, we consider the finite perturbation of the fluid body described by the exponential map with the value of the parameter $\tau=1$. We will drop the subscript $\tau=1$ in the finite values of the functions and denote $\varrho\equiv\varrho_1(\x)$, $K\equiv K_1(\x)$, and ${\bm X}={\bm X}_1(\x)$. The perturbed density of the fluid and gravitational potential of the body at the value $\tau=1$ will be denoted $\mu\equiv\rho_1$ and $\mathfrak{U}\equiv U_1$ respectively. The functions $\rho=\rho(r), p=p(\rho(r))$, and $U=U(r)$ are considered as known and fully determined by the solutions of Eqs. \eqref{3}--\eqref{4}, which define the interior structure of the unperturbed fluid body.

\section{Fundamental Differential Equation for the Height Function}\label{sec-7}

Expanding the left hand-side of Eqs. \eqref{pi6c}, \eqref{pic7} to a Taylor series around the point $\x$ we get the equation of the perturbed 1level surface in the following form:
\bal{ub71}
U(\x)+K(\x)&=&\mathsf{S}_{{\bm X}}^{-1}U(\x)\;,\\
\label{ub72}
\rho(\x)+\varrho(\x)&=&\mathsf{S}_{{\bm X}}^{-1}\rho(\x)\;,
\ea
where the operator $\mathsf{S}_{{\bm X}}=1+\mathsf{T}_{{\bm X}}$, its inverse ${\mathsf S}^{-1}_{{\bm X}}=1+{\mathsf T}^{-1}_{{\bm X}}$, and 
\bal{pa6}
{\mathsf T}_{{\bm X}}&:=&\sum_{n=1}^\infty\frac{1}{n!} X^{i_1}X^{i_2}...X^{i_n}\pd_{i_1i_2...i_n}=\sum_{n=1}^\infty\frac{X^n}{n!}\frac{\pd^n}{\pd r^n}\;,
\ea 
is the operator of translation.
The inverse operator ${\mathsf T}^{-1}_{{\bm X}}$ is defined by condition, ${\mathsf S}^{-1}_{{\bm X}}\cdot{\mathsf S}_{{\bm X}}={\mathsf S}_{{\bm X}}\cdot{\mathsf S}^{-1}_{{\bm X}}=1$. Its solution is the Neumann series:
\bal{bt56}
{\mathsf T}^{-1}_{{\bm X}}&=&\sum_{n=1}^\infty\left(-{\mathsf T}_{{\bm X}}\right)^{n}\;.
\ea
The inverse operator gives the algebraic solution of Eqs. \eqref{ub71}, \eqref{ub72} for the perturbations of gravitational field and density in terms of the height function:
\bal{jk8}
K(\x)&=&\mathsf{T}_{{\bm X}}^{-1}U(\x)\quad,\quad \varrho(\x)=\mathsf{T}_{{\bm X}}^{-1}\rho(\x)\;.
\ea
These equations can be solved algebraically to find the height function $X$ by applying the Lagrange inversion theorem \citep{bellpol} once the perturbation $K$ is determined by solving differential equation \eqref{s14} with appropriate boundary conditions. However, it is more instructive to derive the core differential equation directly for the height function $X$.  

To this end, we start from considering that a partial derivative of the shift operator is:
\bal{xc3}
\pd_i{\mathsf S}_{\bm X}&=&\left(\pd_iX^j\right) {\mathsf S}_{\bm X}\pd_j\;.
\ea
Taking a partial derivative from both sides of Eq. \eqref{ub71} and using Eq. \eqref{xc3} yields
\bal{vv6}
\pd_i U(\x)+\pd_i K(\x)&=&{\mathsf S}^{-1}_{\bm X}M^{-1}_{ij}\pd_jU(\x)\;,
\ea
where the matrix of the deformation gradient
\bal{gt8}
M_{ij}&=&\delta_{ij}+\pd_i X^j\;,
\ea
and $M^{-1}_{ij}$ is its inverse. Notice that $M_{ij}\neq M_{ji}$.
Applying the partial derivative to both sides of Eq. \eqref{vv6} once more yields
\bal{xc9}
\Delta U+\Delta K&=&{\mathsf S}^{-1}_{\bm X}M^{-1}_{ik}\pd_k\left[M^{-1}_{ij}\pd_j U(\x)\right]\;.
\ea
The left hand side of Eq. \eqref{xc9} can be expressed in terms of the density $\rho$ and its perturbation $\varrho$ by using Eq. \eqref{3} and Eq. \eqref{5} for $\tau=1$. After making this replacement, we use Eq. \eqref{ub72} and cancel the shift operator in both sides of the resulting equation. This reduces Eq. \eqref{xc9} to the matrix form of the Poisson-like equation for the height function ${\bm X}=X{\bm n}$:
\bal{me1}
M^{-1}_{ip}(\X)\pd_p\left[M^{-1}_{iq}(\X)\pd_q U(\x)\right]&=&-4\pi\rho(\x)\;.
\ea
This fundamental equation is pivotal in studying the shapes of celestial bodies. It is precise and applicable for determining finite deformations of the body's shape with any required level of rigor.

The inverse matrix $M^{-1}_{ij}$ can be calculated explicitly for the radial diffeomorphism \eqref{pa3}. Calculation is tedious and lengthy and will be given somewhere else. Introducing a new notation $R=r+X$, the result is
\bal{min4}
M^{-1}_{ij}&=&\frac{n^in^j}{R'}+\frac{r}{R}\left[{\cal P}^{ij}+\frac{{\cal P}^{ki}n^j\pd_k R}{R'}\right]\;,
\ea
where ${\cal P}^{ij}=\delta^{ij}-n^in^j$ is the operator of projection onto the plane being orthogonal to the unit vector $n^i$.
Eq. \eqref{min4} allows us to bring the principal equation \eqref{me1} to the more transparent form which we give here for the {\it shape function}: $f=X/r$. After performing several algebraic transformations and laborious calculations, we obtain:
\bal{cre5a}
\Delta f-\frac{2f'}{r}+\frac{6\upbeta}{r}\left(f'+\frac{f}{r}\right)&=&\frac{S(f)}{r}\;.
\ea
Here, the term $S$ in the right hand side of \eqref{cre5a} contains all non-linear terms
\bal{u4iw}
S&\equiv& A'-\upbeta B'+\frac{3\upbeta}{r}(A-B)\;,
\ea
where we have introduced a shorthand notations for two functions
\bal{t5cx}
A&\equiv&\frac{r^2 \pd_if\pd_if}{1+f+rf'}\qquad\quad,\quad\qquad B\equiv 3f^2+f^3\;,
\ea
with the prime denoting a partial derivative with respect to the radial coordinate $r$. Function $\upbeta =\rho/\bar\rho$ is the ratio of the fluid density $\rho=\rho(r)$ to its average value $\bar\rho=\bar\rho(r)$ within the volume of the radius $r$.

Equation \eqref{cre5a} for the shape function $f$ is exact, depending on all three coordinates and exhibiting non-linearity. Remarkably, in the case of constant density, where the parameter ${\upbeta}=1$, some exact solutions to \eqref{cre5a} can be found by substitution:
\bal{xtr5}
f&=&\bigl[1+\chi(\theta,\phi)\bigr]^{-1/2}-1\;,
\ea
which reduces the nonlinear Eq. \eqref{cre5a} to a linear differential equation for function $\chi(\theta,\phi)$. In particular, among these solutions we have found the well-known Maclaurin's ellipsoid of rotation and the Jacobi three-axial ellipsoid \citep{chandr87}.

Unfortunately, equation \eqref{cre5a} cannot be solved analytically in the most general case, necessitating the use of approximations.  The approximate solution can be derived by expanding Eq. \eqref{cre5a} with respect to the shape function $f$ with subsequent using the method of separation of variables, based on the spectral decomposition of $f=f(r,\theta,\phi)$ into spherical harmonics, similar to the canonical theory of figures of celestial bodies \citep{Zharkov-book-1986,horedt_2004book}. The products of the spherical harmonics are decomposed into an irreducible sum over the spherical harmonics with the help of the Clebsch-Gordan coefficients formalism \citep{gelfand_1963}. The boundary conditions for the height function $X$ are derived from the continuity of the normal components of the gravitational field perturbation $K$ and its derivative $K'$ both expressed in terms of $X$ and $X'$ as shown in Eq. \eqref{jk8}.  Proceeding in this way, it is straightforward to verify that Eq. \eqref{cre5a} yields the Clairaut equation \citep{Clairaut-book-1743} in the linearized approximation with respect to the shape function $f$. We have also verified that the quadratic approximation of Eq. \eqref{cre5a} results in the Darwin-de Sitter equations \citep{Darwin_1899,deSitter_1924BAN}.

\section{Celestial Body's Multipole Moments}\label{sec-9}

The perturbed gravitational potential of the body is given by Eq. \eqref{vtf6}. The perturbed density of the body, $\mu=\rho+\varrho$, is expressed in terms of the height function $X$ with the help of Eq. \eqref{jk8}:
\bal{v5d}
\mu(\x)&=&\rho(r)+\mathsf{T}_{{\bm X}}^{-1}\rho(\x)\;.
\ea
Assuming the perturbed fluid body is axially symmetric, the external gravitational field of the body can be expanded in multipoles:  
\bal{cv5}
\mathfrak{U}(\x)=\frac{M}{r}+\sum_{l=2}^\infty\frac{{\cal M}_l}{r^l}P_l(\cos\theta)\;,
\ea
where $M$ is the body' mass, ${\cal M}_l$ are the multipole moments of the body, and $P_l(\cos\theta)$ are the Legendre polynomials. The multipole moments are defined by the volume integral:
\bal{cv6}
{\cal M}_l&=&\int_{\cal V}\mu(\x)r^lP_l(\cos\theta)d^3x\;,
\ea
where the volume element is $d^3x=r^2\sin\theta dr d\theta d\phi$, and the integration is performed over a spherical volume of the unperturbed body with a dimensionless radius $r=1$.

Eq. \eqref{cv6} allows for calculation of the multipole moments of the body with arbitrary accuracy provided that the unperturbed density $\rho$ is known and the height function $X$ has been determined by solving Eq. \eqref{cre5a}. However, the practical calculation of the multipole moments is tedious and and typically performed using approximations. To demonstrate the principle, we perform calculations up to the quadratic order. In this approximation, Eq. \eqref{cv6} reads:
\bal{cv7}
{\cal M}_l&=&\int_{\cal V}\left[-\rho'f+\rho'\left(f^2+rff'\right)+\frac12\rho''f^2r\right]r^{l+3}P_l(\cos\theta)\sin\theta~drd\theta~d\phi\;.
\ea
The shape function $f$ is expanded in Legendre polynomials:
\bal{ss3}
f&=&\sum_{k=0}^\infty f_n(r)P_n (\cos\theta)\;.
\ea
This expansion is substituted into the integrand of Eq. \eqref{cv7} allowing us to calculate the angular integrals. The radial integral can be integrated by parts, discarding the boundary terms since the height function $X$ and all its derivatives vanish at $r=0$, while the density $\rho$ and all its derivatives are zero at the body's boundary for $r=1$. The angular integrals are calculated by making use of the Clebsch-Gordan coefficients formalism. Proceeding in this way we get:
\bal{cv8}
{\cal M}_l&=&4\pi\int_0^1\rho(r)d\left[r^{l+3}F_l(r)\right]\;,\qquad (l\ge 2)
\ea
where functions
\bal{gtr5}
F_l&=&\frac{1}{2l+1}\left[f_l-\left(1+\frac{l}{2}\right)\sum_{n=0}^\infty\sum_{m=0}^\infty \left(C^{nml}\right)^2f_nf_m \right]\;,
\ea
and  
\bal{k1e}
C^{nml}&=&(-1)^{n-m}\sqrt{2l+1}\left(\begin{array}{ccc}
  n & m & l \\
  0& 0 & 0 \\
\end{array}\right)\;,
\ea
are the Clebsch-Gordan coefficients being proportional to the Wigner $3j$ symbols. Expression \eqref{cv8} matches the expression for the multipole moments in Zharkov-Trubitsyn's second order theory \citep{Zharkov-book-1986}.
\section{Conclusions}
This paper addresses the long-standing problem in the theory of figures of astronomical bodies by deriving the fundamental differential equation for finite tidal and rotational deformations. By utilizing the Lie group theory of diffeomorphisms, we bypass the traditional Legendre decomposition of the gravitational potential, a method employed by researchers from A.C. Clairaut to A.M. Lyapunov and V.N. Zharkov. The Lie group approach decouples the differential equations for perturbations in the gravitational field and density, revealing the non-linear nature of Newtonian gravitational interactions within matter. All non-linear terms are then included in the master equation for the height function and can be treated collectively in advanced computer numerical simulations.

Our theory facilitates highly accurate algorithmic calculations of Love numbers, multipole moments, and the internal structure of rapidly rotating giant planets such as Jupiter and Saturn. This advancement not only enhances our understanding of these celestial bodies but also provides a robust framework for future research in planetary science.

Additionally, our findings have significant implications for gravitational wave astronomy. By refining the Newtonian framework for studying the equation of state of matter at nuclear density, our approach offers a more precise tool for analyzing coalescing binary systems, such as neutron star-neutron star or neutron star-black hole pairs. This contributes to a deeper understanding of the dynamics and physical properties of these extreme astrophysical objects, potentially leading to new discoveries in the field of gravitational wave research.

In summary, the innovative application of Lie group theory in this study represents a substantial step forward in both theoretical and computational astrophysics, offering new insights and methodologies for exploring the complex behaviors of astronomical bodies under tidal and rotational influences.

\bibliographystyle{unsrt}
\bibliography{Love_numbers_references}
\end{document}